\documentstyle[prb,preprint,aps]{revtex}
\draft

\begin{document}
\title{Spin Glass Behavior in the Ru-1222 and Ru-1212 rutheno-cuprate families: a
comparative study.}
\author{C. A. Cardoso$^{1,\ast }$, F. M. Araujo-Moreira$^{1}$, V. P. S. Awana$^{2}$,
H. Kishan$^{2}$, E. Takayama-Muromachi$^{3}$, O. F. de Lima$^{4}$}
\address{$^{1}$Grupo de Materiais e Dispositivos, CMDMC, Departamento de F\'{i}sica,\\
UFSCar, Caixa Postal 676, 13565-905 , S\~{a}o Carlos-SP, Brazil\\
$^{2}$National Physical Laboratory, K. S. Krishnan Marg., New Delhi 110012,\\
India\\
$^{3}$Superconducting Materials Center, National Institute for Materials\\
Science, 1-1 Namiki, Tsukuba, Ibaraki 305-0044, Japan\\
$^{4}$Instituto de F\'{i}sica ''Gleb Wataghin'', UNICAMP, 13083-970,\\
Campinas-SP, Brazil}
\date{April. 10, 2003}
\maketitle

\begin{abstract}
Strong evidences of a spin glass phase on polycrystalline $%
RuSr_{2}Ln_{1.5}Ce_{0.5}Cu_{2}O_{10-\delta }$ (LnRu-1222, Ln = Y, Ho and Dy)
is provided by ac susceptibility ($\chi _{ac}$) and dc magnetization
measurements. The fingerprint of a spin glass transition, the
frequency-dependent cusp in $\chi _{ac}$ vs. $T$ measurements at low
magnetic field, is observed for all LnRu-1222 samples. The change in the
cusp position with frequency follows the Vogel-Fulcher law, which is
commonly accepted to describe a spin glass with magnetically interacting
clusters. The strong supression of the cusp in the ac susceptibility
measurement in the presence of magnetic fields as low as 500 $Oe$, and the
results of thermoremanent magnetization (TRM) and isothermal remanent
magnetization (IRM) measurements are also evidences of a spin glass
behavior. By comparing these results with measurements on a YRu-1212 sample
we conclude that the glassy behavior is characteristic of the Ru-1222 phase,
not being observed for the studied YRu-1212 sample. We suggest that the spin
glass transition occur only in the Ru-1222 phase possibly due to the larger
number of oxygen vacancies present in these compounds.
\end{abstract}

\pacs{75.50.Lk, 75.40.Gb, 75.60.Ej, }

\section{Introduction}

A large number of reports have recently appeared focusing on the magnetic
and superconducting properties of the rutheno-cuprates $%
RSr_{2}Ln_{2-x}Ce_{x}Cu_{2}O_{10-\delta }$ (LnRu-1222) \cite
{Felner97,Felner98,Felner2002PRB65,Hirai02,Jardim02,Awana02,Knee00} and $%
RuSr_{2}LnCu_{2}O_{8-\delta }$ (LnRu-1212) \cite
{Awana02b,Chen01,Bernhard99,Tallon99,Pingle99,Fainstein99,Lynn00,Williams,Tokunaga,Takagiwa,Chmaissem,Jorgensen,Butera}%
. However, despite the intensive research on these materials, some
unanswered questions still remain. For instance, the oxygen
non-stoichiometry, carrier concentration and valence state of Ru are not
well understood yet \cite{reviewAwana,Matvejeff}. One of the most
controversial questions is the exact type of magnetic ordering in the
LnRu-1222 family. In contrast to the LnRu-1212 family, for which a consensus
has been reached on the canted antiferromagnetic ordering for the Ru
sublattice \cite{Lynn00,Butera}, the detailed magnetic ordering of the
LnRu-1222 family is still lacking. Although the magnetic behavior of
LnRu-1222 has been considered to be analogous to the magnetic response for
LnRu-1212 samples, some recent results point towards various differences
between them \cite{Felner2002PRB65,FelnerLA,cardoso03}. In particular,
strong evidences of spin glass behavior was found  in GdRu-1222 \cite
{cardoso03}. In the present work we extend this previous study to other
LnRu-1222 compositions (Ln = Y, Dy, Ho). The verification of the spin glass
behavior for these new samples and the differences between these results
with those obtained for an YRu-1212 sample lead us to conclude that the
spin-glass phase is characteristic of the Ru-1222 phase. We propose that the
presence of oxygen vacancies in the RuO$_{6}$ octahedra for Ru-1222 may
cause the frustration of the magnetic ordering of the Ru ions, leading to a
glassy behavior in this compounds, which does not usually occur for Ru-1212
samples. Studies comparing the oxygen non-stoichiometry for both families
support such interpretation \cite{reviewAwana,Matvejeff}.

\section{Experimental Details}

Samples of composition $RuSr_{2}Ln_{1.5}Ce_{0.5}Cu_{2}O_{10-\delta }$
(LnRu-1222) with Ln = Y, Ho, and Dy were synthesized through a high-pressure
high-temperature (HPHT) solid-state reaction route \cite{Awanasintese}.
Briefly, a mixture of $%
RuO_{2}+SrO_{2}+SrCuO_{2}+3/4CuO+1/4CuO_{0.011}+3/4Ln_{2}O_{3}+1/2CeO_{2}$
was sealed in a gold capsule and submitted to a pressure of 6 GPa in a
flat-belt-type-high-pressure apparatus, where it was allowed to react for 2
hours at 1200 $%
%TCIMACRO{\UNICODE[m]{0xb0}}%
%BeginExpansion
{{}^\circ}%
%EndExpansion
C$. It is expected that this procedure provides samples with oxygen content
close to nominal \cite{Awanasintese}, i. e. 10. All LnRu-1222 samples
present a tetragonal structure within the {\it I4/mmm} space group as
confirmed by X-ray powder diffraction (XRD) patterns obtained at room
temperature (Philips-PW1800; CuK$_{\alpha }$ radiation). The lattice
parameters are $a=b=3.824(1),\;3.819(1),$ and $3.813(1)$ \AA\ and $%
c=28.445(1),\;28.439(1),$ and$\;28.419(1)$ \AA\ for Ln = Dy, Y and Ho,
respectively. The YRu-1212 was prepared using the same procedure, except
that for this compound it was necessary to start from a slightly Ru poor
composition in order to obtain a single phase sample of the desired
stoichiometry \cite{Awana03,Muromachi01}. The lattice parameter obtained are 
$a=b=3.818(1)$ \AA\ and $c=11.5222(3)$ \AA . The details of the precise HPHT
process are given in Ref. [27]. All samples are single phase within the XRD
resolution, except for YRu-1222 which presents a small amount of $SrRuO_{3}$ 
\cite{Awanasintese}. All ac susceptibility measurements were performed in a
commercial PPMS (Physical Properties Measurement System), while for the dc
measurements a SQUID magnetometer MPMS-5 were employed, both equipments made
by Quantum Design company.

\section{Experimental Results}

The ac susceptibility $(\chi _{ac}=\chi ^{\prime }+i\chi ^{\prime \prime })$
technique is a very powerful method to provide evidence of a spin-glass
behavior. In this case, both components $\chi ^{\prime }$ and $\chi ^{\prime
\prime }$ of $\chi _{ac}$ present a sharp, frequency dependent cusp. The
position of the cusp in $\chi ^{\prime }$ defines the freezing temperature $%
T_{f}$, which is coincident with the temperature of the inflection point in $%
\chi ^{\prime \prime }$. It is also well known that dc magnetic fields as
low as a few hundreds of Oersted may suppress this cusp. In Fig. 1 we
present the ac susceptibility for all four studied samples measured at a
frequency $\nu =1000$ $Hz$ for three different dc fields $H=50,$ $150$ and $%
500$ Oe. All samples present a well defined peak at temperatures $%
T_{f}=87.42,$ $103.3$ and $101.5$ $K$ for Dy-, Ho- and YRu-1222, and $%
T_{N}=140.6$ $K$ for YRu-1212. All three LnRu-1222 samples present a strong
suppression \ of the peak observed in $\chi ^{\prime }$, which almost
disappeared for $H=500$ $Oe$. Besides this suppression, the peak position is
shifted to higher temperatures at higher fields. On the other hand, the peak
for YRu-1212 present a much weaker dependence with the applied field and it
is slightly shifted to lower temperatures with the increase of $H$. Another
important difference between LnRu-1222 samples and the YRu-1212 one is the
presence of an anomaly in the susceptibility at $T_{M}=150$ $K$ for
LnRu-1222. Sample YRu-1212 does not present any anomaly in $\chi ^{\prime }$
above $T_{p}$. This anomaly is more prominent for Y- and $\ $HoRu-1222 than
for DyRu-1222, and is smeared out when the applied field is increased. It is
striking that the anomaly observed in LnRu-1222 samples occurs at the same
temperature, independently of Ln, and also that this temperature is close to
the position of the peak observed for YRu-1212. These results are the first
evidence that the peaks observed in $\chi ^{\prime }$ have different origins
for Ru-1222 and Ru-1212 families. In Fig. 2 we present the dc susceptibility
as a function of temperature for different magnetic fields. The curves for
all LnRu-1222 show a similar behavior. The field cooled (FC) measurements
present a ferromagnetic-like shape, while the zero-field cooled (ZFC) curve
\ present a well defined peak at temperatures, for $H=50$ $Oe$, $T_{p}=83.7,$
$101.8$ and $140.8$ K for Dy-, Ho- and YRu-1222 respectively. The two branch
apart twice, one at $T_{M}\approx 150$ $K$ (for all LnRu-1222 samples), and
again at $T_{irr}\approx T_{f}\sim 100$ $K$ (Y- and HoRu-1222) or $84$ $K$
(DyRu-1222). At $H=50$ $Oe$ the ZFC and FC curves are not exactly reversible
for any temperature below $T_{M}$, so $T_{irr}$ is not well determined. As
the applied field is increased to 500 $Oe$, a significant reduction in the
irreversibility is observed and both, ZFC and FC curves, tend to a
ferromagnetic-like behavior. The irreversibility observed at temperatures
above the peak in the ZFC curve is also greatly affected by the increase of
the applied field, being hardly distinguishable at $H=500$ $Oe$. Comparing
the results for dc magnetization with the ac susceptibility, we observe: (1)
the anomaly in $\chi ^{\prime }$ occurs at the same temperature of the
deviation from the paramagnetic behavior in the magnetization data (at $T_{M}
$); (2) the peak position in $\chi ^{\prime }$ roughly coincides with $%
T_{irr}$. The coincidence of the peak temperature in $\chi ^{\prime }$ with $%
T_{irr}$, the drastic reduction of the irreversibility and the
ferromagnetic-like behavior of both ZFC and FC curves with quite small
applied fields, are all consistent with the expected behavior of a
spin-glass system. Now we turn our attention to the YRu-1212 sample again.
The ZFC/FC irreversibility is also suppressed with the increase of $H$, but
a significant irreversibility remains present even at $H=500$ $Oe$. Also,
the peak in the ZFC curve is broadened at higher $H$, but it do not
disappear (up to 500 Oe). Once more, the behavior of the YRu-1212 is quite
different from the observed for LnRu-1222, which in turn is very similar
with each other.

The most clear and conclusive way to experimentally separate an
antiferromagnet from a spin-glass is to probe the frequency dependence of
the peak in $\chi ^{\prime }.$ As can be observed in Fig. 3, the peak shifts
to lower temperatures and its intensity increases as the frequency of the
excitation field is decreased, for all three LnRu-1222 samples. This
frequency dependence of $\chi ^{\prime }$ is a typical feature of the
dynamics of spin-glass systems \cite{Mydoshbook}. On the other hand, no
significant changes in the peak are observed for the YRu-1212 sample, which
is the expected behavior expected for an antiferromagnet. A quantitative
measure of the frequency shift is obtained from $\Delta T_{f}/[T_{f}\log
(\omega )]$. This quantity varies in the range of 0.004 - 0.018 for
spin-glass systems \cite{Mydoshbook}, while for superparamagnets \cite
{Mydoshbook} it is of the order of 0.3. From the FC susceptibility
measurements at different frequencies, presented in Fig. 3, we could
estimate $\Delta T_{f}/[T_{f}\log (\omega )]\approx 0.0031,$ 0.0021 and ,
0.0018 respectively for $Dy-$, $Ho-$ and $YRu-1222$, which is slightly low
but yet consistent with expected behavior for spin-glasses.

There are basically two different possible interpretations of the spin-glass
freezing: the first one assumes the existence of a true equilibrium phase
transition at a finite temperature (canonical spin glasses \cite{Edwards}).
The second interpretation assumes the existence of magnetic clusters and, in
this case, the freezing is a nonequilibrium phenomenon \cite{Tholence}. For
isolated clusters (superparamagnets), the frequency dependence of their
freezing temperature (in this context more correctly referred as {\it %
blocking temperature}) has been predicted to follow an Arrhenius law 
\begin{equation}
\omega =\omega _{0}\exp [-E_{a}/k_{B}T_{f}],  \label{Arrhenius}
\end{equation}
where $E_{a}$ is the potential barrier which separates two easy orientations
of the cluster and $\omega $ is the driving frequency of the $\chi _{ac}$
measurement. However, for magnetically interacting clusters, a Vogel-Fulcher
law has been proposed, which has the form:

\begin{equation}
\omega =\omega _{0}\exp [-E_{a}/k_{B}(T_{f}-T_{0})],  \label{VF law}
\end{equation}
where $T_{0}$ can be viewed as a phenomenological parameter which describes
the intercluster interactions. Equation \ref{VF law} implies a linear
dependence of the freezing temperature with $1/\ln [(\omega \tau
_{0})^{-1}], $ $\tau _{0}=1/\nu _{0}=2\pi /\omega _{0}.$ In Fig. 4 we
present Vogel-Fulcher plots for the three LnRu-1222 samples, which shows
that our data follows the expected linear behavior. From the best linear fit
we obtained $\nu _{0}\approx 1\times 10^{12}$ $Hz$, $T_{0}=83.14,$ 99.83, $%
98.80 $ $K$ and $E_{a}=90.25,$ 74.35, 58.38 $K$ for $Dy-$, $Ho-$ and $%
YRu-1222$, respectively.

Also, the existence of the spin-glass behavior has been checked through the
measurement of its remanent magnetization. The remanent state can be
achieved by two different procedures. The first one consists on cooling the
sample in a certain magnetic field H from $T>>T_{f}$ down to the measurement
temperature. After the temperature is stabilized the applied field is
removed and the measurement is performed. This procedure gives us the
thermoremanent magnetization (TRM). In the spin-glass case, the TRM quickly
reaches its saturation value for low values of H. The second way to measure
the remanence is to cool the sample in zero field down to the measurement
temperature, and only after that a magnetic field H is turned on. After a
short time the field is removed (the temperature remains constant) and the
measurement is performed. This provides the isothermal remanent
magnetization (IRM), which saturates at a higher H than the observed for the
TRM. Both TRM and IRM coincide in the saturated region \cite{Mydoshbook}.
Fig. 5 shows both TRM and IRM measurements for the three LnRu-1222 and the
YRu-1212 samples. For the TRM experiments, all\ curves present an abrupt
increase at lower fields, going through a small maximum just before reaching
its saturation, while the IRM experiments, all curves present a much more
gentle increase up to the saturation value. These results are in agreement
with the expected behavior of a spin-glass, as exposed above. The TRM and
IRM results for the YRu-1212 sample present a less pronounced dependence on
the applied field, although it also increases with $H$.

\section{Discussion}

The results presented in the previous section strongly support that all
LnRu-1222 studied samples present a spin-glass transition at $T_{f}$, while
the YRu-1212 sample is an antiferromagnet with spin canting. Once the
magnetic behavior of the studied samples has been established, we have now
to compare the different results obtained for each sample. This is done in
two steps. First we compare the results for the LnRu-1222 to check for
possible influences of the specific Ln ion on the spin-glass properties of
the sample. A second and more important discussion is to find out why
Ru-1222 and Ru-1212 present so distinct magnetic properties. In order to
start these discussions, it is useful to review the crystallographic
structure of both systems.

The structure of $RuSr_{2}LnCu_{2}O_{8-\delta }$ (LnRu-1212) is derived from
that of $REBa_{2}Cu_{3}O_{7-\delta }$ (usually known as RE-123; in a more
general categorization scheme \cite{Karppinen}, it is written as $%
CuBa_{2}RECu_{2}O_{7-\delta },\;Cu-1212$). To go from the former to the last
structure, it is necessary to replace the $Cu$ in the charge reservoir by $%
Ru,$ thus transforming the $CuO$ chains in $RuO_{2}$ planes, and the $Ba$
ions are substituted by $Sr$. The structure of $%
RuSr_{2}(Ln,Ce)_{2}Cu_{2}O_{10-\delta }$ (LnRu-1222), on the other hand,
results from that of LnRu-1212 when a three-layer fluorite-type block $%
(Ln,Ce)-O_{2}-(Ln,Ce)$ substitutes the oxygen-free $Ln$ layer, between the
two adjacent $CuO_{2}$ planes. It is believed that in rutheno-cuprates the $%
RuO_{6}$ octahedra in the charge reservoir are mainly responsible for both
the magnetic ordering and the hole doping of the superconductive $CuO_{2}$
planes, while the $Ln$ ions do not order magnetically. The paramagnetic
contribution of magnetic $Ln$ ions is usually associated to the increase of
the magnetization observed at low temperatures in these systems. Then it is
clear that the $Ln$ layer plays a less important role in determining the
magnetic ordering in these ruthenates. In fact, it could be inferred from
the data presented in Figs. 1 - 5 that the change of the $Ln$ ion did not
change qualitatively the magnetic behavior of the LnRu-1222 samples.
However, a few differences between them can be pointed, the more obvious of
them being the variation of the freezing temperature as one goes from Ho to
Dy or Gd \cite{cardoso03}. As shown in Fig. 6, this is related to the
changes of the volume of the unit cell, due to the different ionic radii of
the $Ln$ ions. In the same figure is also plotted $T_{M}$ which is almost
constant for the samples studied in this work but, for the GdRu-1222 sample
previously studied, the value of $T_{M}$ is much higher \cite{cardoso03}.
Remembering that the GdRu-1222 sample was prepared by a different technique 
\cite{AwanaSint2}, it is more likely that this difference is a consequence
of the synthesis process than an intrinsic effect of the different $Ln$
ions. Also, it is observed that the irreversibility observed in ZFC/FC
magnetization curves at temperatures $T_{f}<T<T_{M}$ is quite small for
GdRu-1222 and becomes more prominent as smaller $Ln$ ions are considered.
Both, different ionic radii and synthesis process, seem to be important in
the determination of this irreversibility.

A more intriguing question to be answered is why all Ru-1222 samples we have
studied present a clear spin-glass behavior while the YRu-1212 sample is an
antiferromagnet. This is intriguing because the differences between both
structures are restricted just to the $Ln$ plane, which is isolated from the
RuO$_{2}$ planes which order magnetically, as already discussed. Therefore,
it seems probable that the $Ln$ plane would affect the magnetic behavior of
the compound by indirectly affecting the RuO$_{2}$ planes. In order to
produce a spin-glass state it is necessary to have a frustration of the
magnetic order, which is accomplished by disorder. One could first try to
find this randomness in the mixing of $Ce$ and $Ln$, which is only present
in the Ru-1222 family. A random distribution of $Ln$ could cause small local
perturbations in the crystalline structure in such a way that the RuO$_{6}$
could be distorted, thus inducing the frustration of Ru magnetic moments. A
more likely and simple explanation would come from oxygen non-stoichiometry,
which could directly affect the RuO$_{2}$ planes. The different magnetic
behavior of the two families (Ru-1212 and Ru-1222) could only be explained
if the Ru-1222 system are more susceptible to present oxygen vacancies than
the Ru-1212 family. In fact, a work recently reported \cite{Matvejeff}
concludes that GdRu-1212 can be obtained with a near stoichiometric oxygen
content, while the GdRu-1222 phase is clearly oxygen-deficient even after
100-atm O$_{2}$ annealing. Changes observed in the valence of Ru in
GdRu-1222 as the oxygen content varies suggest that the vacancies occur in
the RuO$_{2}$ layers. The same work also shown that the oxygen content in
GdRu-1212 is almost constant upon various annealings, while for GdRu-1222 a
wider range in oxygen content is accessible. These results lead us to
conclude that the most reasonable origin for the glassy behavior observed in
our LnRu-1222 samples is the presence of oxygen vacancies. If this is true,
our results show in a quite dramatic way the influence of the exact oxygen
content of the sample. It is likely that some of the contradictory results
reported on Ru-1222 in the literature are consequences of the difficulty to
fully oxygenate samples of the Ru-1222 system. Although we do not observed
any indication of a spin glass phase in the studied YRu-1212, we believe
that degraded, oxygen-depleted Ru-1212 may also present a glassy behavior.
It is important to notice that the samples we have studied in this work (as
well as the GdRu-1222 sample, reported in Ref. [24]) were synthesized by
routes which allow a higher oxygen content than more usual procedures.

\section{Concluding Remarks}

In summary, in this work we show that the frequency-dependent peak observed
in the temperature dependence of the ac susceptibility $\chi _{ac}$, its
suppression by a small magnetic field, combined with the remanent
magnetization results, provide strong evidence of the existence of a spin
glass phase over a significant temperature range in polycrystalline
LnRu-1222. This is valid for samples with different $Ln$ and prepared by two
different routes (considering also the GdRu-1222 sample, previously reported 
\cite{cardoso03}), indicating that the glassy behavior is a characteristic
feature of the Ru-1222 family. This is to be contrasted with the existence
of long-range antiferromagnetic order with spin canting for the YRu-1212
sample. We believe that oxygen vacancies could frustrate the long-range
order of the Ru spins, leading to a spin glass phase. Recent results
pointing that the Ru-1222 phase is usually oxygen deficient, while the
Ru-1212 family can be synthesized with near stoichiometric oxygen content,
corroborate this idea and provide a possible explanation for the different
behavior we found for these two families.

This work was supported by Brazilian agencies FAPESP and CNPq.

\label{aReferencias}

\vspace{0.5cm}

FIG. 1. Real part of the ac susceptibility as a function of temperature for
different magnetic fields, for all four samples. Insets show the anomaly
observed (in LnRu-1222 samples only) at $T_{f}<T<T_{M}.$

\vspace{0.5cm}

FIG. 2. dc magnetization as a function of temperature for different magnetic
fields, for all four samples. Insets show the irreversibility observed (in
LnRu-1222 samples only) at $T_{f}<T<T_{M}.$

\bigskip

FIG. 3. Real part of the ac susceptibility as a function of temperature for $%
H=50$ Oe and four different frequencies, for all four samples. The peak
position defines the freezing temperature $T_{f}$ (for the $LnRu-1222$
samples)$.$

\vspace{0.5cm}

FIG. 4. Variation of the freezing temperature $T_{f}$ with the frequency of
the ac field in a Vogel-Fulcher plot, for the three LnRu-1222 samples. The
solid line is the best fit of Eq. \ref{VF law}.

\vspace{0.5cm}

FIG. 5. Thermoremanent magnetization (TRM, solid symbols) and isothermal
remanent magnetization (IRM, open symbols) for all samples, measured as a
function of the applied field, at $T=30$ K.

\vspace{0.5cm}

FIG. 6. Variation of the freezing temperature $T_{f}$ and of the magnetic
transition $T_{M}$ as a function of the volume of the unit cells for
LnRu-1222 compounds. The results for GdRu-1222 were extracted from Ref. [24].

\end{document}